\documentstyle[pra,aps,psfig]{revtex}
\def\Im{\mathop{\rm Im}} 
\def\Tr{\mathop{\rm Tr}}

\begin{document}
\title{Fluctuations and dissipation for a mirror in vacuum}
\author{Marc Thierry Jaekel $^{(a)}$ and Serge Reynaud $^{(b)}$}
\address{(a) Laboratoire de Physique Th\'{e}orique de l'ENS
\thanks{%
Unit\'e propre du Centre National de la Recherche Scientifique,
associ\'ee \`a l'Ecole Normale Sup\'erieure et \`a l'Universit\'e
Paris-Sud}, 24 rue Lhomond, F75231 Paris Cedex 05 France\\
(b) Laboratoire de Spectroscopie Hertzienne
\thanks{%
Unit\'e de l'Ecole Normale Sup\'erieure et de l'Universit\'e
Pierre et Marie Curie, associ\'ee au Centre National de la Recherche 
Scientifique}, 4 place Jussieu, case 74, F75252 Paris Cedex 05 France}
\date{{\sc Quantum Optics} {\bf 4} (1992) 39-53}
\maketitle

\begin{abstract}
A mirror in the vacuum is submitted to a radiation pressure exerted by
scattered fields. It is known that the resulting mean force is zero for a
motionless mirror, but not for a mirror moving with a non-uniform
acceleration. We show here that this force results from a motional
modification of the field scattering while being associated with the
fluctuations of the radiation pressure on a motionless mirror. We consider
the case of a scalar field in a two-dimensional spacetime and characterize
the scattering upon the mirror by frequency dependent transmissivity and
reflectivity functions obeying unitarity, causality and high frequency
transparency conditions. We derive causal expressions for dissipation and
fluctuations and exhibit their relation for any stationary input. We recover
the known damping force at the limit of a perfect mirror in vacuum. Finally,
we interpret the force as a mechanical signature of the squeezing effect
associated with the mirror's motion.
\end{abstract}

\section*{Introduction}

Even in the vacuum state, the electromagnetic field exhibits quantum
fluctuations \cite{Fluctuat1} which manifest themselves through the
macroscopic Casimir forces \cite{Fluctuat2,Fluctuat3}.

These forces can be understood as resulting from the radiation pressure
exerted by the scattered fluctuations and they depend upon the reflection
coefficients which characterize the boundaries. Assuming that the boundaries
are transparent at high frequencies, which is certainly the case for any
real mirrors, one obtains expressions free from the divergences usually
associated with the infiniteness of the vacuum energy \cite{Fluctuat4}.

In this formulation of the Casimir effect, the force is related to the
vacuum stress tensor evaluated on the boundaries and is itself a fluctuating
quantity. As illustrated by the Langevin theory of Brownian motion \cite
{Fluctuat5}, any fluctuating force has a long term cumulative effect. Here a
motional force for a mirror in vacuum can be deduced from linear response
theory \cite{Fluctuat6} and it is connected to the fluctuations through some
`fluctuation-dissipation relations'. A force has yet been derived for a
perfectly reflecting mirror moving in a two-dimensional (2D) spacetime \cite
{Fluctuat7,Fluctuat8}; it is dissipative and proportional to the third time
derivative of the mirror's position $q$ (in a linear approximation with
respect to $q$) 
\begin{equation}
F\left( t\right) =\frac{\hbar q^{\prime \prime \prime }\left( t\right) }{%
6\pi c^ 2 }  \eqnum 1 
\end{equation}
(from now on, we use natural units where $c=1$; however, we keep $\hbar $ as
a scale for vacuum fluctuations). This force results from a motional
modification of the vacuum stress tensor and is connected to the Casimir
forces. Actually, both effects are present when the motion of two mirrors is
studied \cite{Fluctuat8,Fluctuat9,Fluctuat10}. However, the expression (1)
of the force does not exhibit the causal properties which are expected from
the linear response theory.

A related effect has been studied in great detail since it limits the
sensitivity of the interferometers designed for gravitational wave detection 
\cite{Fluctuat11,Fluctuat12,Fluctuat13,Fluctuat14}. When irradiated by a
laser wave, a mirror undergoes a fluctuating radiation pressure \cite
{Fluctuat15} as well as a damping force proportional to its velocity and to
the laser intensity \cite{Fluctuat16}. However, the discussion of these
effects has not taken into account the fact that they remain at the limit of
a null laser intensity; the radiation pressure fluctuates also in the vacuum
and this causes an extra mirror's damping.

In the present paper, we study the simplest case where a point like mirror
is placed in the stationary state of a scalar field in a 2D spacetime (with
the vacuum as a particular case). The field scattering upon the mirror is
characterized by frequency dependent transmissivity and reflectivity
functions obeying unitarity, causality and high frequency transparency
conditions.

First, we derive the radiation pressure exerted upon a motionless mirror.
Then, we study the motional modification of the field scattering (at first
order in the mirror's displacement) and obtain a causal expression for the
motional force. We exhibit the relation connecting this force with the
fluctuations of the radiation pressure computed for a motionless mirror.
These results are demonstrated for any stationary state of the input fields.
At the end of the paper, we give the particular expressions for the vacuum
state. Equation (1) is reproduced at frequencies well below the reflection
cutoff. Finally, the motional force is connected with the squeezing of
vacuum field, as put into evidence by the expression of the effective
Hamiltonian describing the mirror's motion in the linear approximation.

\section*{Notations}

In a 2D spacetime (time coordinate $t$, space coordinate $x$), a free scalar
field is the sum of two counterpropagating components $\varphi (t-x)+\psi
(t+x)$; we will write these two components in a column matrix 
\[
\Phi _{x}(t)=\left( 
\begin{array}{c}
\varphi (t-x) \\ 
\psi (t+x)
\end{array}
\right) 
\]
We will consider that any function $f$ defined in the time domain and its
Fourier transforms $f$ are related through \footnote{{\it The notation used
in the original paper for Fourier transforms has been changed to a more 
convenient one.}} 
\[
f(t)=\int \frac{{\rm d}\omega }{2\pi }f[\omega ]e^{-i\omega t} 
\]
For example, the Fourier development of the column $\Phi _{x}$ is related to
the standard annihilation and creation operators corresponding to the two
propagation directions 
\begin{eqnarray}
&&\Phi _{x}[\omega ] =\left( 
\begin{array}{c}
\varphi [\omega ]e^{i\omega x} \\ 
\psi [\omega ]e^{-i\omega x}
\end{array}
\right) =e^{i\eta \omega x}\Phi [\omega ]\qquad \eta =\left( 
\begin{array}{cc}
1 & 0 \\ 
0 & -1
\end{array}
\right)  \nonumber \\
&&\varphi [\omega ] =\sqrt{\frac{\hbar }{2|\omega |}}\left( \theta (\omega
)a_\omega +\theta (-\omega )a_{-\omega }^\dagger \right) \qquad \psi
[\omega ] =\sqrt{\frac{\hbar }{2|\omega |}}\left( \theta (\omega )b_{\omega
}+\theta (-\omega )b_{-\omega }^\dagger \right)  \eqnum 2 
\end{eqnarray}
The abbreviated notation $\Phi $ is used for the value of $\Phi _{x}$
evaluated at $x=0$. The commutation relations of the Fourier components of
the fields are 
\begin{equation}
\left[ \varphi [\omega ],\varphi [\omega ^\prime ]\right] =\left[ \psi
[\omega ],\psi [\omega ^\prime ]\right] =2\pi \delta (\omega +\omega
^\prime )\frac{\hbar }{2\omega }\qquad \left[ \varphi [\omega ],\psi
[\omega ^\prime ]\right] =0  \eqnum{3}
\end{equation}

We will use specific notations for the fields $\overline{\Phi }$ evaluated
at the time dependent mirror's position $q_{t}$ (shortened notation for 
$q(t) $), written as a function of the mirror's proper time $\tau $, as well
as for its Fourier transforms 
\[
\overline{\Phi }(\tau )=\Phi _{q_{t}}(t)=\left\{ e^{-x\eta \partial
_{t}}\Phi (t)\right\} _{x=q_{t}}\qquad {\rm d}\tau =\sqrt{1-q_{t}^{\prime \
2}}{\rm d}t\qquad \overline{\Phi }(\tau )=\int \frac{{\rm d}\omega }{2\pi }
\overline{\Phi }[\omega ]e^{-i\omega \tau } 
\]
As $\overline{\Phi }$ and $\Phi $ are related through a phase modulation,
there is no simple relation between their Fourier transforms, except in the
particular case of a motionless or uniformly moving mirror. In order to deal
with this transformation, we will perform a first order expansion in a
modification $\delta q_{t}$ of the mirror's trajectory around $q=0$ 
\begin{equation}
\overline{\Phi }(t)=\Phi (t)-\delta q_{t}\eta \partial _{t}\Phi (t) 
\eqnum{4a}
\end{equation}
As second order terms are neglected, the proper time $\tau $ and the
laboratory time $t$ coincide. Equivalently, in the frequency domain 
\begin{equation}
\overline{\Phi }[\omega ]={\int }\frac{{\rm d}\omega ^\prime }{2\pi }
\left( 2\pi \delta (\omega -\omega ^\prime )+\delta q[\omega -\omega
^\prime ]\eta i\omega ^\prime \right) \Phi [\omega ^\prime ] 
\eqnum{4b}
\end{equation}

The energy and impulsion densities correspond to two counterpropagating
energy fluxes 
\[
e_{x}(t)=\varphi ^\prime (t-x)^ 2 +\psi ^\prime (t+x)^ 2 \qquad
p_{x}(t)=\varphi ^\prime (t-x)^ 2 -\psi ^\prime (t+x)^ 2  
\]
Their mean values may be written in terms of the covariance matrix, the
elements of which are the two point correlation functions of the fields 
\begin{eqnarray*}
&&C_{x,x^\prime }(t,t^\prime )=\left\langle \Phi _{x}(t)\Phi _{x^{\prime
}}(t^\prime )^{\rm T}\right\rangle \\
\left\langle e_{x}(t)\right\rangle &=&\left\{ \Tr\left[ \partial
_{t}\partial _{t^\prime }C_{x,x^\prime }(t,t^\prime )\right] \right\}
_{t^\prime =t}\qquad \left\langle p_{x}(t)\right\rangle =\left\{ \Tr
\left[ \eta \partial _{t}\partial _{t^\prime }C_{x,x^{\prime
}}(t,t^\prime )\right] \right\} _{t^\prime =t}
\end{eqnarray*}
$\Tr$ stands for the trace operation on square matrices and $X^{\rm T}
$ for the transposed of $X$. The same expressions written in the frequency
domain will be useful, particularly 
\begin{eqnarray}
&&C_{x,x^\prime }[\omega ,\omega ^\prime ]=\left\langle \Phi _{x}[\omega
]\Phi _{x^\prime }[\omega ^\prime ]^{\rm T}\right\rangle =e^{i\eta
\omega x}C[\omega ,\omega ^\prime ]e^{i\eta \omega ^\prime x^\prime } 
\nonumber \\
&&\left\langle e_{x}(t)\right\rangle =\int \frac{{\rm d}\omega }{2\pi }\int 
\frac{{\rm d}\omega ^\prime }{2\pi }e^{-i\omega t-i\omega ^{\prime
}t}i\omega i\omega ^\prime \Tr\left[ C_{x,x}[\omega ,\omega ^{\prime
}]\right]  \eqnum{5} \\
&&\left\langle p_{x}(t)\right\rangle =\int \frac{{\rm d}\omega }{2\pi }\int 
\frac{{\rm d}\omega ^\prime }{2\pi }e^{-i\omega t-i\omega ^{\prime
}t}i\omega i\omega ^\prime \Tr\left[ \eta C_{x,x}[\omega ,\omega
^\prime ]\right]  \nonumber 
\end{eqnarray}

For a stationary state, the covariance matrices depend only upon one
parameter 
\begin{equation}
C(t,t^\prime )=c(t-t^\prime )\qquad C[\omega ,\omega ^\prime ]=2\pi
\delta (\omega +\omega ^\prime )c[\omega ]  \eqnum{6}
\end{equation}
We will often write the covariances in terms of the anticommutators which
characterize the various states of the fields and of the commutators which
do not depend upon the state (see equation 3) 
\begin{eqnarray}
&&c[\omega ] = c_{+}[\omega ]+c_{-}[\omega ]  \eqnum{7} \\
c_{+}[\omega ] &=& \frac{c[\omega ]+c(-\omega )^{\rm T}} 2 =c_{+}(-\omega
)^{\rm T} \qquad c_{-}[\omega ] =\frac{c[\omega ]-c(-\omega )^{\rm T}} 2 
=\frac{I\hbar }{4\omega }  \nonumber
\end{eqnarray}
$I$ is the unit matrix.

\section*{Scattering upon a motionless mirror}

In the limiting case of perfect reflection, the field is constrained to be
zero at the mirror's position $q$ so that the input and output fields (see
Figure 1) are related through 
\begin{eqnarray*}
\psi _{\rm out}(t+q) &=&-\varphi _{\rm in}(t-q)\qquad \varphi _{\rm out}
(t-q)=-\psi _{\rm in}(t+q) \\
\Phi _{\rm out}[\omega ] &=&e^{-i\eta \omega q}\left( 
\begin{array}{cc}
0 & -1 \\ 
-1 & 0
\end{array}
\right) e^{i\eta \omega q}\Phi _{\rm in}[\omega ]
\end{eqnarray*}

For a partly transmitting mirror, the scattering of the field is described
by a frequency dependent $S-$matrix 
\begin{eqnarray}
\Phi _{\rm out}[\omega ] &=&e^{-i\eta \omega q}\overline{S}[\omega
]e^{i\eta \omega q}\Phi _{\rm in}[\omega ] \qquad \overline{S}[\omega ]
=\left( 
\begin{array}{cc}
\overline{s}[\omega ] & \overline{r}[\omega ] \\ 
\overline{r}[\omega ] & \overline{s}[\omega ]
\end{array}
\right)  \eqnum{8}
\end{eqnarray}
For clarity, we denote $\overline{S}$ the $S-$matrix in the proper frame
(same convention as for the fields). As a consequence of the translational
invariance of a stationary state, all the results will be independent of $q$
and we shall suppose from now on that $q=0$.

\begin{figure}[h]
\centerline{\psfig{figure=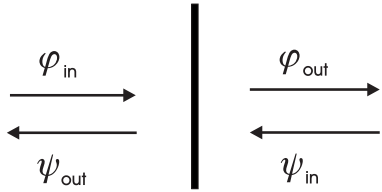,width=5cm}}
\caption{The mirror scatters the two counterpropagating fields.}
\label{Fig1}
\end{figure}

The matrix $\overline{S}$ is supposed to obey the following conditions \cite
{Fluctuat4}: it is real in the temporal domain, causal and unitary 
($\overline{S}=\overline{S}^{\rm T}$) 
\begin{eqnarray*}
&&\overline{S}[-\omega ]=\overline{S}[\omega ]^{*} \\
&&\overline{S}[\omega ]{\rm \ is\ analytic\ and\ regular\ for\ }
\Im \omega >0 \\
&&\overline{S}[\omega ]\overline{S}[\omega ]^\dagger =1
\end{eqnarray*}
Finally, the mirror is supposed transparent at high frequencies 
\begin{equation}
\overline{S}[\omega ]\rightarrow I{\rm \quad for\ }\omega \rightarrow \infty
\eqnum{9}
\end{equation}
This assumption will allow regularization of the ultraviolet divergences
associated with the infiniteness of the vacuum energy. It must be noted that
the perfect mirror ($s=0$ and $r=-1$ at all frequencies) does not obey this
condition. So, it will be preferable to consider the perfect mirror as the
limit of a model obeying the transparency condition (for example a mirror
perfectly reflecting at frequencies below a reflection cutoff).

\section*{Mean radiation pressure upon a motionless mirror}

The force $F(t)$ may be evaluated as the difference between the radiation
pressures exerted upon the left and right sides of the mirror at rest at 
$q=0 $. In a 2D spacetime, the component $T_{xx}$ of the stress tensor is
equal to the energy density and one gets 
\[
F(t)=\varphi _{\rm in}^{\prime \ 2}(t)+\psi _{\rm out}^{\prime \
2}(t)-\varphi _{\rm out}^{\prime \ 2}(t)-\psi _{\rm in}^{\prime \ 2}(t) 
\]
This force can also be considered as the difference between the impulsion
densities of the input and output fields evaluated at the mirror's position 
\begin{equation}
F(t)=p_{\rm in}(t)-p_{\rm out}(t)  \eqnum{10}
\end{equation}

For a perfect mirror, the force is twice the impulsion density which would
exist at the location of the mirror in its absence \cite{Fluctuat9}. The
mean value of this force is zero in the vacuum state. However, we shall see
later on that the instantaneous radiation pressure has irreducible quantum
fluctuations.

For a partly transmitting mirror, the force $F(t)$ is still given by the
difference (10) between the input and output impulsion densities but we have
now to evaluate the output fields by using the input output relation (8) 
\[
C_{\rm out}[\omega ,\omega ^\prime ]=\overline{S}[\omega ]C_{\rm in}
[\omega ,\omega ^\prime ]\overline{S}[\omega ^\prime ] 
\]
One gets an expression of the force having the same form as equation (5) 
\[
\left\langle F(t)\right\rangle =\int \frac{{\rm d}\omega }{2\pi }\int \frac{
{\rm d}\omega ^\prime }{2\pi }e^{-i\omega t-i\omega ^\prime t}i\omega
i\omega ^\prime \Tr\left[ {\cal F}[\omega ,\omega ^\prime ]C_{\rm in}
[\omega ,\omega ^\prime ]\right] 
\]
where ${\cal F}$ is a square matrix 
\begin{eqnarray}
{\cal F}[\omega ,\omega ^\prime ] &=&\eta -\overline{S}[\omega ^{\prime
}]\eta \overline{S}[\omega ]=\left( 
\begin{array}{cc}
\alpha [\omega ,\omega ^\prime ] & \beta [\omega ,\omega ^\prime ] \\ 
-\beta [\omega ,\omega ^\prime ] & -\alpha [\omega ,\omega ^\prime ]
\end{array}
\right)  \nonumber \\
\alpha [\omega ,\omega ^\prime ] &=&1-\overline{s}[\omega ]\overline{s}
[\omega ^\prime ]+\overline{r}[\omega ]\overline{r}[\omega ^\prime ] 
\nonumber \\
\beta [\omega ,\omega ^\prime ] &=&\overline{s}[\omega ]\overline{r}
[\omega ^\prime ]-\overline{r}[\omega ]\overline{s}[\omega ^\prime ] 
\eqnum{11}
\end{eqnarray}

The matrix ${\cal F}$ obeys the following properties which will be used
thereafter 
\begin{eqnarray}
{\cal F}[\omega ,\omega ^\prime ]^{\rm T} &=&{\cal F}[\omega ^{\prime
},\omega ]=\eta {\cal F}[\omega ,\omega ^\prime ]\eta  \eqnum{12a} \\
{\cal F}[\omega ,\omega ^\prime ]^\dagger  &=&{\cal F}[-\omega ^{\prime
},-\omega ]  \eqnum{12b}
\end{eqnarray}
\begin{eqnarray}
{\cal F}[\omega ,\omega ^\prime ]{\cal F}[\omega ,\omega ^{\prime
}]^\dagger  &=&{\cal F}[\omega ,\omega ^\prime ]\eta +\eta {\cal F}
[\omega ,\omega ^\prime ]^\dagger   \eqnum{13a} \\
{\cal F}[\omega ,\omega ^\prime ]^\dagger {\cal F}[\omega ,\omega
^\prime ] &=&\eta {\cal F}[\omega ,\omega ^\prime ]+{\cal F}[\omega
,\omega ^\prime ]^\dagger \eta  \eqnum{13b}
\end{eqnarray}
Using equations (7), the force is written in terms of the field
anticommutators 
\begin{equation}
\left\langle F(t)\right\rangle =\int \frac{{\rm d}\omega }{2\pi }\int \frac{
{\rm d}\omega ^\prime }{2\pi }e^{-i\omega t-i\omega ^\prime t}i\omega
i\omega ^\prime \Tr\left[ {\cal F}[\omega ,\omega ^\prime ]C_{+,
{\rm in}}[\omega ,\omega ^\prime ]\right]  \eqnum{14}
\end{equation}

The force may be written in the temporal domain 
\[
\left\langle F(t)\right\rangle =\left\{ \partial _{t}\partial _{t^\prime }
\Tr \left[ \eta C_{\rm in}(t,t^\prime )-\eta C_{\rm out}
(t,t^\prime )\right] \right\} _{t^\prime =t} 
\]
with 
\begin{eqnarray}
\Tr\left[ \eta C_{\rm in}(t,t^\prime )-\eta C_{\rm out}
(t,t^\prime )\right] &=&{\int }{\rm d}t^{\prime \prime }{\int }{\rm d}
t^{\prime \prime \prime }\Tr\left[ {\cal F}(t^{\prime \prime
},t^{\prime \prime \prime })C_{\rm in}(t-t^{\prime \prime },t^{\prime
}-t^{\prime \prime \prime })\right]  \nonumber \\
{\cal F}(t,t^\prime ) &=&\delta (t^\prime )\eta \delta (t)-\overline{S}
(t^\prime )\eta \overline{S}(t)  \eqnum{15}
\end{eqnarray}
It clearly appears on these expressions that the force exerted upon the
mirror is a retarded function of the input stress tensor: $\overline{S}$ is
a causal function and ${\cal F}(t,t^\prime )$ is zero as soon as $t<0$ or 
$t^\prime <0$.

The expression (14) provides the mean force for any input state. For a
stationary input (see equation 6), one gets a simpler expression 
\[
\left\langle F\right\rangle =\int \frac{{\rm d}\omega }{2\pi }\omega ^ 2 
\Tr \left[ {\cal F}(\omega ,-\omega )c_{+,{\rm in}}[\omega ]\right] 
\]

One can evaluate also the energy exchange between the field and the mirror;
it is the difference between the energy densities of the input and output
fields 
\[
G(t)=e_{\rm in}(t)-e_{\rm out}(t) 
\]
One finds that $G$ is given by equation (14) with ${\cal F}$ replaced by 
\[
{\cal G}[\omega ,\omega ^\prime ]=I-\overline{S}[\omega ^\prime ]
\overline{S}[\omega ] 
\]
As $\overline{S}$ is unitary 
\[
{\cal G}(\omega ,-\omega )=0 
\]
Therefore, the energy exchange is zero in any stationary state 
\begin{equation}
\left\langle G\right\rangle =\int \frac{{\rm d}\omega }{2\pi }\omega ^ 2 
\Tr \left[ {\cal G}(\omega ,-\omega )c_{\rm in}[\omega ]\right] =0 
\eqnum{16}
\end{equation}

\section*{Scattering upon a moving mirror}

At perfect reflection, the field is still zero on a moving mirror \cite
{Fluctuat9} and the input output relations have a simple form for the fields
evaluated along the mirror's trajectory 
\begin{eqnarray*}
\psi _{\rm out}(t+q_{t})=-\varphi _{\rm in}(t-q_{t}) &\qquad &\varphi _
{\rm out}(t-q_{t})=-\psi _{\rm in}(t+q_{t}) \\
\overline{\psi }_{\rm out}(\tau )=-\overline{\varphi }_{\rm in}(\tau )
&\qquad &\overline{\varphi }_{\rm out}(\tau )=-\overline{\psi }_{\rm in}
(\tau ) \\
\overline{\Phi }_{\rm out}[\omega ] &=&\left( 
\begin{array}{cc}
0 & -1 \\ 
-1 & 0
\end{array}
\right) \overline{\Phi }_{\rm in}[\omega ]
\end{eqnarray*}
These relations describe the Doppler shift associated with the mirror's
motion (Lorentz transformation for the frequencies) and the dilatation of
the derived fields $\varphi ^\prime $ and $\psi ^\prime $ (Lorentz
transformation for the fields) \cite{Fluctuat9}.

For a partly transmitting mirror, the $S-$matrix (defined previously for a
motionless mirror) describes the scattering of the field evaluated along the
trajectory \footnote{%
This assumption could be justified by considering a conformal transformation
from the laboratory to a `proper frame' \cite{Fluctuat8} which preserves the
two counterpropagating components and is chosen so that the mirror is at
rest and the time is the mirror's proper time in the proper frame.} 
\[
\overline{\Phi }_{\rm out}[\omega ]=\overline{S}[\omega ]\overline{\Phi }_
{\rm in}[\omega ] 
\]
The $S-$matrix is deduced in a first order expansion in a mirror's
displacement $\delta q_{t}$ by using the transformation (4) 
\begin{eqnarray}
&&\Phi _{\rm out}[\omega ]={\int }\frac{{\rm d}\omega ^\prime }{2\pi }
\left( 2\pi \delta (\omega -\omega ^\prime )\overline{S}[\omega ]+\delta
S[\omega ,\omega ^\prime ]\right) \Phi _{\rm in}[\omega ^\prime ] 
\nonumber \\
&&\delta S[\omega ,\omega ^\prime ]=i\omega ^\prime \delta q[\omega
-\omega ^\prime ]\left( \overline{S}[\omega ]\eta -\eta \overline{S}
[\omega ^\prime ]\right)  \eqnum{17}
\end{eqnarray}

\section*{Force exerted upon a moving mirror}

Taking the mirror's motion into account, the force can be written as 
\[
\left\langle F(t)\right\rangle =\left\langle p_{q_{t},{\rm in}}(t)-p_{q_{t},
{\rm out}}(t)\right\rangle -q_{t}^\prime \left\langle e_{q_{t},{\rm in}
}(t)-e_{q_{t},{\rm out}}(t)\right\rangle 
\]
where the densities are evaluated at the mirror's position. In a first order
expansion in $\delta q_{t}$, it becomes 
\begin{eqnarray*}
\left\langle F(t)\right\rangle &=&\left\langle p_{\rm in}(t)-p_{\rm out}
(t)\right\rangle -\delta q_{t}\partial _{t}\left\langle G(t)\right\rangle
-\delta q_{t}^\prime \left\langle G(t)\right\rangle \\
\left\langle G(t)\right\rangle &=&\left\langle e_{\rm in}(t)-e_{\rm out}
(t)\right\rangle
\end{eqnarray*}
where the densities are evaluated at $x=0$, the second term represents the
variation of the impulsion densities in a translation (for a free field 
$\partial _{x}p+\partial _{t}e=0$) and the third term is the correction
proportional to energy densities and to the mirror's velocity. The energy
modification $\left\langle G(t)\right\rangle $ has to be evaluated at the
zeroth order. From now on, we will restrict ourselves to stationary inputs,
in which case $\left\langle G(t)\right\rangle $ is zero (see equation 16).
The two corrections associated with it will be forgotten in the expression
of $\left\langle F(t)\right\rangle $.

We will eventually compute the mean force as 
\[
\left\langle \delta F(t)\right\rangle =-\left\langle \delta p_{\rm out}
(t)\right\rangle =-\left\{ \partial _{t}\partial _{t^\prime } \Tr
\left[ \eta \delta C_{\rm out}(t,t^\prime )\right] \right\}
_{t=t^\prime } 
\]
where the variations of the output fields are due to the modification 
$\delta S$ of the $S-$matrix in the laboratory 
\begin{eqnarray*}
\left\langle \delta F(t)\right\rangle &=&\int \frac{{\rm d}\omega }{2\pi }
\int \frac{{\rm d}\omega ^\prime }{2\pi }e^{-i\omega t-i\omega ^{\prime
}t}\omega \omega ^\prime \Tr\left[ \eta \delta C_{\rm out}[\omega
,\omega ^\prime ]\right] \\
\delta C_{\rm out}[\omega ,\omega ^\prime ] &=&\int \frac{{\rm d}\omega
^{\prime \prime }}{2\pi }\left( \delta S[\omega ,\omega ^{\prime \prime }]C_
{\rm in}[\omega ^{\prime \prime },\omega ^\prime ]\overline{S}[\omega
^\prime ]^{\rm T}+\overline{S}[\omega ]C_{\rm in}[\omega ,\omega
^{\prime \prime }]\delta S[\omega ^\prime ,\omega ^{\prime \prime }]^
{\rm T}\right)
\end{eqnarray*}
that is, for a stationary input (see equations 6 and 17) 
\begin{eqnarray}
\delta C_{\rm out}[\omega ,\omega ^\prime ] &=&-i\omega ^\prime \delta
q[\omega +\omega ^\prime ]\left( \overline{S}[\omega ]\eta -\eta \overline{S}
[-\omega ^\prime ]\right) c_{\rm in}[-\omega ^\prime ]\overline{S}
[\omega ^\prime ]  \nonumber \\
&&-i\omega \delta q[\omega +\omega ^\prime ]\overline{S}[\omega ]c_{\rm in}
[\omega ]\left( \eta \overline{S}[\omega ^\prime ]-\overline{S}[-\omega
]\eta \right)  \eqnum{18}
\end{eqnarray}
Using the unitarity of $\overline{S}$, one obtains 
\[
\Tr\left[ \eta \delta C_{\rm out}[\omega ,\omega ^\prime ]\right]
=\delta q[\omega +\omega ^\prime ]\Tr\left[ {\cal F}[\omega ,\omega
^\prime ]\left( i\omega c_{\rm in}[\omega ]\eta +i\omega ^\prime \eta
c_{\rm in}[-\omega ^\prime ]\right) \right] 
\]

We will write the motional force as a symmetric integral over the two
frequencies 
\begin{eqnarray*}
\left\langle \delta F(t)\right\rangle  &=&\int \frac{{\rm d}\omega }{2\pi }
\int \frac{{\rm d}\omega ^\prime }{2\pi }e^{-i\omega t-i\omega ^{\prime
}t}\chi [\omega ,\omega ^\prime ]\delta q[\omega +\omega ^\prime ] \\
\chi [\omega ,\omega ^\prime ] &=&\frac{\omega \omega ^\prime } 2 \Tr
\left[ {\cal F}[\omega ,\omega ^\prime ]\left( i\omega c_{\rm in}
[\omega ]\eta +i\omega ^\prime \eta c_{\rm in}[-\omega ^{\prime
}]\right) +{\cal F}[\omega ^\prime ,\omega ]\left( i\omega ^\prime c_
{\rm in}[\omega ^\prime ]\eta +i\omega \eta c_{\rm in}[-\omega ]\right)
\right] 
\end{eqnarray*}
Transposing the matrices inside the second line and using the properties
(12), one transforms $\chi [\omega ,\omega ^\prime ]$ into a function of
the anticommutators (see equation 7) 
\begin{eqnarray}
\chi [\omega ,\omega ^\prime ] &=&\omega \omega ^\prime \Tr\left[ 
{\cal F}[\omega ,\omega ^\prime ]\left( i\omega c_{+,{\rm in}}[\omega
]\eta +i\omega ^\prime \eta c_{+,{\rm in}}[-\omega ^\prime ]\right)
\right]   \nonumber \\
&=&\omega \omega ^\prime \Tr\left[ {\cal F}[\omega ,\omega ^{\prime
}]i\omega c_{+,{\rm in}}[\omega ]\eta +{\cal F}[\omega ^\prime ,\omega
]i\omega ^\prime c_{+,{\rm in}}[\omega ^\prime ]\eta \right]   \eqnum{19}
\end{eqnarray}

Finally, the force appears as a linear response to the mirror's motion 
\begin{equation}
\left\langle \delta F[\omega ]\right\rangle =\chi [\omega ]\delta q[\omega ]
\eqnum{20}
\end{equation}
with a susceptibility given by 
\begin{equation}
\chi [\omega ]=\int \frac{{\rm d}\omega ^\prime }{2\pi }\chi [\omega
^\prime ,\omega -\omega ^\prime ]  \eqnum{21}
\end{equation}
These expressions generalize the motional force (1) known for a perfect
mirror in the vacuum to the case of a partly transmitting mirror in an
arbitrary stationary input field. Later on, we shall see that the force (1)
is recovered as an approximate result.

We have shown that the motional force is a consequence of the transformation
of the fields by the moving mirror. Actually, this transformation is a
squeezing effect: the particular case where the input state is the vacuum is
discussed later on. In order to squeeze the field, the mirror has to
exchange energy with it. A motional force thus appears as the signature of
the squeezing effect.

\section*{Interpretation of the motional force in the comoving frame}

It is instructive to compute the force in the comoving frame 
\[
\left\langle \overline{F}(\tau )\right\rangle =\left\langle \overline{p}_
{\rm in}(\tau )-\overline{p}_{\rm out}(\tau )\right\rangle =\left\{
\partial _{\tau }\partial _{\tau ^\prime }\Tr\left[ \eta \overline{C}
_{\rm in}(\tau ,\tau ^\prime )-\eta \overline{C}_{\rm out}(\tau ,\tau
^\prime )\right] \right\} _{\tau =\tau ^\prime } 
\]
This expression differs from the force computed in the laboratory but the
corrections are seen to depend upon the energy exchange $\left\langle 
\overline{G}(\tau )\right\rangle $ and they can be forgotten (see the
previous discussion). The mean force is the same in the laboratory or in the
comoving frame in a first order expansion in $\delta q_{t}$.

In the comoving frame, one can use the expression of the force computed for
a motionless mirror but the input stress tensor has to be modified because
of the mirror's motion 
\[
\left\langle \delta \overline{F}(t)\right\rangle =\int \frac{{\rm d}\omega }
{2\pi }\int \frac{{\rm d}\omega ^\prime }{2\pi }e^{-i\omega t-i\omega
^\prime t}i\omega i\omega ^\prime \Tr\left[ {\cal F}[\omega ,\omega
^\prime ]\delta \overline{C}_{\rm in}[\omega ,\omega ^\prime ]\right] 
\]
The apparent stress tensor is obtained from the transformation (4) of the
fields 
\[
\delta \overline{C}_{\rm in}(t,t^\prime )=-\delta q_{t}\eta \partial _{t}
\overline{C}_{\rm in}(t,t^\prime )-\partial _{t^\prime }\overline{C}_
{\rm in}(t,t^\prime )\eta \delta q_{t^\prime } 
\]
that is, for a stationary input 
\begin{equation}
\delta \overline{C}_{\rm in}[\omega ,\omega ^\prime ]=\delta q[\omega
+\omega ^\prime ]\left( -i\omega c_{\rm in}[\omega ]\eta -i\omega
^\prime \eta c_{\rm in}[-\omega ^\prime ]\right)  \eqnum{22}
\end{equation}
This gives exactly the same force as previously.

So the force exerted upon a moving mirror can be computed in the laboratory
by considering that the $S-$matrix is modified (see equation 17) or in the
comoving frame by considering that the input stress tensor is modified (see
equation 22). It clearly appears in the comoving frame that the force is a
causal function of the mirror's trajectory (see equation 15). We will
discuss this point more precisely in the particular case of a vacuum input.

\section*{Fluctuations of the radiation pressure upon a motionless mirror}

The appearance of a force for a moving mirror can actually be guessed by
inspecting the situation where the mirror is at rest. Indeed, we shall now
exhibit the quantitative relation between the motional force and the noise
spectrum of the force computed for a motionless mirror. This connection can
be considered as a `fluctuation-dissipation' theorem for a mirror which
scatters field fluctuations. We will thus check in this context that the
motional force can be deduced from linear response theory.

As a qualitative introduction to the problem of force fluctuations, we
consider the expression (10) which relates the force and the impulsion
densities $p_{\rm in}$ and $p_{\rm out}$ of the input and output fields.
We see that the instantaneous impulsion density has irreducible quantum
fluctuations. For example the two counterpropagating energy fluxes are
statistically independent random variables in the vacuum state and the force
fluctuations do not vanish.

We come now to a quantitative evaluation of the correlation function $C_{FF}$
of the force exerted upon a partly transmitting mirror 
\[
C_{FF}(t)=\left\langle F(t)F(0)\right\rangle -\left\langle F\right\rangle
^ 2  
\]
Using the input-output relation (8), one obtains an operatorial expression
of the force analogous to equation (14) 
\begin{equation}
F(t)=\int \frac{{\rm d}\omega }{2\pi }\int \frac{{\rm d}\omega ^\prime }
{2\pi }e^{-i\omega t-i\omega ^\prime t}i\omega i\omega ^\prime \Tr
\left[ {\cal F}[\omega ,\omega ^\prime ]\Phi _{\rm in}[\omega ]\Phi _
{\rm in}[\omega ^\prime ]^{\rm T}\right]  \eqnum{23}
\end{equation}
It follows that $C_{FF}$ depends upon four-points correlation functions of
the fields. Inserting the expressions (2) of the input fields in terms of
the annihilation and creation operators, we compute the four-points
functions ($\Phi _{\alpha }$ stands for a component $\varphi $ or $\psi $ of
the input field; we suppose that the input field is in a stationary Gaussian
state; this is the case for the vacuum; this property is equivalent to the
Wick's rules \cite{Fluctuat17}) 
\begin{eqnarray*}
&&\left\langle \Phi _{\alpha }[\omega ]\Phi _{\beta }[\omega ^\prime ]\Phi
_{\alpha ^\prime }[\omega ^{\prime \prime }]\Phi _{\beta ^\prime
}[\omega ^{\prime \prime \prime }]\right\rangle -\left\langle \Phi _{\alpha
}[\omega ]\Phi _{\beta }[\omega ^\prime ]\right\rangle \left\langle \Phi
_{\alpha ^\prime }[\omega ^{\prime \prime }]\Phi _{\beta ^\prime
}[\omega ^{\prime \prime \prime }]\right\rangle \\
&&\qquad =2\pi \delta (\omega +\omega ^{\prime \prime })c_{\alpha \alpha
^\prime }[\omega ]2\pi \delta (\omega ^\prime +\omega ^{\prime \prime
\prime })c_{\beta \beta ^\prime }[\omega ^\prime ]+2\pi \delta (\omega
+\omega ^{\prime \prime \prime })c_{\alpha \beta ^\prime }[\omega ]2\pi
\delta (\omega ^\prime +\omega ^{\prime \prime })c_{\beta \alpha ^\prime
}[\omega ^\prime ]
\end{eqnarray*}
The autocorrelation function of the force thus comes out as (using the
properties 12) 
\begin{eqnarray}
C_{FF}(t) &=&\int \frac{{\rm d}\omega }{2\pi }\int \frac{{\rm d}\omega
^\prime }{2\pi }e^{-i\omega t-i\omega ^\prime t}C_{FF}[\omega ,\omega
^\prime ]  \eqnum{24} \\
C_{FF}[\omega ,\omega ^\prime ] &=&2\omega ^ 2 \omega ^{\prime \ 2}\Tr 
\left[ {\cal F}[\omega ,\omega ^\prime ]c_{\rm in}[\omega ]{\cal F} 
[\omega ,\omega ^\prime ]^\dagger c_{\rm in}[\omega ^\prime ]^{\rm T}
\right]  \eqnum{25}
\end{eqnarray}
It appears that $C_{FF}$ is a symmetric function of the two frequencies 
\[
C_{FF}[\omega ,\omega ^\prime ]=C_{FF}[\omega ^\prime ,\omega ] 
\]

One obtains the noise spectrum of the force as the Fourier transforms of the
autocorrelation function (24) 
\begin{equation}
C_{FF}[\omega ]=\int \frac{{\rm d}\omega ^\prime }{2\pi }C_{FF}[\omega
^\prime ,\omega -\omega ^\prime ]  \eqnum{26}
\end{equation}
The explicit evaluation of this noise spectrum for a mirror in the vacuum
will be given later on.

\section*{Commutator of the force operator}

In order to exhibit the fluctuation-dissipation relations, we have to
compute the mean value of the commutator of the force operator 
\[
\xi _{FF}(t)=\frac{C_{FF}(t)-C_{FF}(-t)}{2\hbar }
\]
We write it 
\begin{eqnarray}
&&\xi _{FF}(t)=\int \frac{{\rm d}\omega }{2\pi }\int \frac{{\rm d}\omega
^\prime }{2\pi }e^{-i\omega t-i\omega ^\prime t}\xi _{FF}[\omega ,\omega
^\prime ]  \eqnum{27} \\
&&\xi _{FF}[\omega ,\omega ^\prime ]=\frac{C_{FF}[\omega ,\omega ^\prime
]-C_{FF}[-\omega ^\prime ,-\omega ]}{2\hbar }  \eqnum{28}
\end{eqnarray}
Using the properties (12), one transforms this expression into 
\[
\xi _{FF}[\omega ,\omega ^\prime ]=\frac{\omega ^ 2 \omega ^{\prime \ 2}}
{\hbar }\Tr\left[ {\cal F}[\omega ,\omega ^\prime ]c_{\rm in}[\omega
]{\cal F}[\omega ,\omega ^\prime ]^\dagger c_{\rm in}[\omega ^\prime
]^{\rm T}-{\cal F}[\omega ,\omega ^\prime ]c_{\rm in}[-\omega ]^{\rm T}
{\cal F}[\omega ,\omega ^\prime ]^\dagger c_{\rm in}[-\omega
^\prime ]\right] 
\]
One then writes the covariances in terms of anticommutators and commutators
(see equations 7) 
\[
\xi _{FF}[\omega ,\omega ^\prime ]=\frac{\omega \omega ^\prime } 2 \Tr
\left[ {\cal F}[\omega ,\omega ^\prime ]\omega c_{+,{\rm in}}[\omega ]
{\cal F}[\omega ,\omega ^\prime ]^\dagger +{\cal F}[\omega ,\omega
^\prime ]{\cal F}[\omega ,\omega ^\prime ]^\dagger \omega ^\prime
c_{+,{\rm in}}[\omega ^\prime ]^{\rm T}\right] 
\]
We could have derived this expression directly from the field commutators.
Therefore, it is correct even when the input fields are not Gaussian
variables.

Finally, one uses the property (13) to obtain (compare with the expression
19) 
\begin{equation}
\xi _{FF}[\omega ,\omega ^\prime ]=\frac{\chi [\omega ,\omega ^\prime
]-\chi [-\omega ,-\omega ^\prime ]}{2i}  \eqnum{29}
\end{equation}
A comparison with equation (28) shows that there is a connection between the
force fluctuations computed for a motionless mirror and the mean force
computed for a moving mirror.

\section*{Fluctuation-dissipation relation}

Defining the Fourier transforms of the force commutator (27) 
\begin{equation}
\xi _{FF}[\omega ]=\int \frac{{\rm d}\omega ^\prime }{2\pi }\xi
_{FF}[\omega ^\prime ,\omega -\omega ^\prime ]  \eqnum{30}
\end{equation}
one deduces from equations (28) and (29) 
\begin{equation}
\xi _{FF}[\omega ]=\frac{C_{FF}[\omega ]-C_{FF}[-\omega ]}{2\hbar }=\frac{
\chi [\omega ]-\chi [-\omega ]}{2i}  \eqnum{31}
\end{equation}
This constitutes the fluctuation-dissipation relation for a mirror submitted
to the radiation pressure of scattered field fluctuations.

This establishes that the motional force can be derived from the correlation
functions using the linear response theory. Considering that a classical
modification $\delta q_{t}$ of the mirror's trajectory corresponds to an
effective perturbation of the Hamiltonian 
\begin{equation}
\delta H(t)=-F(t)\delta q_{t}  \eqnum{32}
\end{equation}
where $F(t)$ is the force operator, we get the motional force (20) by using
the formulas of linear response theory given in the appendix. The
susceptibility $\chi [\omega ]$ is actually the retarded response function
while the function $\xi _{FF}[\omega ]$ is the spectral density associated
with the effective perturbation (32).

The fluctuation-dissipation relation (31) provides the spectral density $\xi
_{FF}$ as soon as the noise spectrum or the susceptibility is known. Using
the analytic properties of the response functions, the latter can therefore
be deduced from the noise spectrum. The force for a moving mirror can always
be guessed by inspecting the fluctuations of the force upon a motionless
mirror.

The converse is not true in general. However, if the input state corresponds
to a thermal equilibrium, the anticommutator of the force (the fluctuations)
can be obtained from the commutator (the susceptibility). The vacuum state
is the equilibrium state at zero temperature and the noise spectrum $C_{FF}$
can effectively be deduced in this case from the spectral density $\xi _{FF}$, 
as we see now.

\section*{The case of vacuum fluctuations}

The covariance matrix corresponding to the vacuum state is easily derived
from the expressions (2) of the fields in terms of the annihilation and
creation operators 
\[
c_{\rm vac}(\omega )=I\theta (\omega )\frac{\hbar }{2\omega }\qquad
c_{+,{\rm vac}}(\omega )=I\frac{\hbar }{4\left| \omega \right| } 
\]
This covariance matrix is scalar so that the mean radiation pressure (14) is
zero as expected.

The function $\chi $ (see equation 19) corresponding to the vacuum is 
\[
\chi _{\rm vac}[\omega ,\omega ^\prime ]=\frac{i\hbar } 2 \omega \omega
^\prime \left( \varepsilon (\omega )+\varepsilon (\omega ^{\prime
})\right) \alpha [\omega ,\omega ^\prime ] 
\]
where $\varepsilon $ is the sign function 
\[
\varepsilon (\omega )=\theta (\omega )-\theta (-\omega ) 
\]
and $\alpha $ the diagonal term of the matrix ${\cal F}$ (see equation 11).
It follows that the susceptibility for a mirror which scatters the vacuum
fluctuations may be written (see equation 21) 
\begin{equation}
\chi _{\rm vac}[\omega ]=i\hbar \int_{0}^\omega \frac{{\rm d}\omega ^\prime 
}{2\pi }\omega ^\prime (\omega -\omega ^\prime )\alpha [\omega ^\prime,
\omega -\omega ^\prime ]  \eqnum{33}
\end{equation}

If the frequency $\omega $ is such that $s=0$ and $r=-1$ between $0$ and 
$\omega $, $\alpha $ may be replaced by $2$ in equation (33) and the response
function reduces to 
\begin{equation}
\chi _{\rm vac}[\omega ]=i\hbar \frac{\omega ^{3}}{6\pi }  \eqnum{34}
\end{equation}
This means that the damping force in the vacuum can be approximated by
equation (1), in the limiting case of a perfect mirror, at frequencies below
the reflection cutoff. In other words, the coarse grained force (force
averaged over a time longer than the reflection delay) is proportional to 
$q^{\prime \prime \prime }$. However, it has to be emphasized that the exact
expression (33) is causal which is not the case for the approximated one
(34). Regularized and causal expressions are obtained only when considering
partly transmitting mirrors \cite{Fluctuat4}.

The force (1) is identical to the linear approximation (first order
expansion in the mirror's displacement $\delta q$) of the non linear
expression obtained by Fulling and Davies for a perfectly reflecting mirror 
\cite{Fluctuat8}. It is worth to note that it is also the non relativistic
limit ($q^\prime \ll c=1$) of this expression. This suggests that the
domain of validity of the first-order expansion corresponds to a
non-relativistic mirror velocity.

The vacuum fields may be considered as Gaussian random variables \cite
{Fluctuat17} so that we can effectively compute the force correlations from
the field covariance matrix. One gets from equations (25) and (13) 
\[
C_{FF,{\rm vac}}[\omega ,\omega ^\prime ]=\hbar ^ 2 \theta (\omega )\theta
(\omega ^\prime )\omega \omega ^\prime \left( \alpha [\omega ,\omega
^\prime ]+\alpha [\omega ,\omega ^\prime ]^{*}\right) 
\]
One then deduces the noise spectrum of the force 
\begin{equation}
C_{FF,{\rm vac}}[\omega ]=2\hbar \theta (\omega )\xi _{FF,{\rm vac}}[\omega ]\qquad \xi
_{FF,{\rm vac}}[\omega ]=\frac{\chi _{\rm vac}[\omega ]-\chi _{\rm vac}[-\omega ]}{2i} 
\eqnum{35}
\end{equation}

Clearly, the fluctuation-dissipation relation is obeyed in the vacuum state.
Simple results are obtained when the reflection is perfect at frequencies
between 0 and $\omega $ 
\begin{eqnarray}
C_{FF,{\rm vac}}[\omega ] &=&\frac{\hbar ^ 2 }{3\pi }\theta (\omega )\omega ^{3} 
\eqnum{36} \\
\xi _{FF,{\rm vac}}[\omega ] &=&\frac{\hbar }{6\pi }\omega ^{3}  \eqnum{37}
\end{eqnarray}
The connection between the variation of $C_{FF}$ as $\omega ^{3}$ and the
variation of the damping force as $q^{\prime \prime \prime }$ is a
manifestation of the fluctuation-dissipation relation. As already noted, the
noise spectrum (35) is more regular at high frequencies than the
approximation (36) as a consequence of the transparency condition (9).

Two properties of the noise spectrum (35) have to be emphasized. First, 
$C_{FF}$ is zero at the limit of a null frequency, which means that the force
fluctuations are averaged to zero when integrated over a long time. Note
that the input impulsion density $p_{\rm in}$ also vanishes when
integrated over a long time.

Second, the relation (35) between the noise spectrum $C_{FF}$ and the
spectral density $\xi _{FF}$ implies that the noise spectrum contains only
positive frequency components, because the vacuum is the zero temperature
state. It follows that the vacuum can damp the mirror's motion but cannot
excite it.

\section*{Connection with squeezing}

We have developed a formalism where the scattering is characterized by
frequency dependent coefficients. The effect of the mirror's motion is
described by a modification of the $S-$matrix in the laboratory or by a
transformation of the input stress tensor in the comoving frame. It can be
noted that the scattering formalism has already been used for dealing with
vacuum fluctuations in accelerated frames or in curved space \cite
{Fluctuat18,Fluctuat19,Fluctuat20,Fluctuat21,Fluctuat22,Fluctuat23}.

In this scattering approach, the damping force for a mirror in the vacuum
appears as connected to squeezing \cite{Fluctuat24,Fluctuat25}. In order to
put this point into evidence, we write the secular part (component at zero
frequency) of the effective Hamiltonian (32) as follows (see the operatorial
expression 23 of the force) 
\[
\delta H[0]={\int }{\rm d}t\ \delta H(t)=\int \frac{{\rm d}\omega }{2\pi }
\int \frac{{\rm d}\omega ^\prime }{2\pi }\delta q[-\omega -\omega ^\prime
]\omega \omega ^\prime \Tr\left[ {\cal F}[\omega ,\omega ^\prime
]\Phi _{\rm in}[\omega ]\Phi _{\rm in}[\omega ^\prime ]^{\rm T}
\right] 
\]
Considering as an example the case where the mirror oscillates at a fixed
frequency 2$\omega _{0}$ 
\[
\delta q_{t}=\delta q_{0}\cos (2\omega _{0}t) 
\]
one recognizes an effective Hamiltonian giving rise to a squeezing effect 
\cite{Fluctuat1}.

Actually, the motional force constitutes a mechanical consequence of the
squeezing. As the input field is the vacuum state, the mirror has to give
energy to the field in order to squeeze it and this damps its motion.

In the laboratory frame, the modification of the output covariance matrix
which results from the effective Hamiltonian is given by equation (18) 
\[
\delta C_{\rm out}[\omega ,\omega ^\prime ]=\frac{i\hbar } 2 \delta
q[\omega +\omega ^\prime ]\left( \theta (\omega )-\theta (-\omega ^\prime
)\right) {\cal F}[\omega ^\prime ,\omega ] 
\]
For the oscillating mirror, this is non zero only if the two frequencies 
$\omega $ and $\omega ^\prime $ have the same sign and if their sum is $\pm
2\omega _{0}$. It follows that the squeezing effect vanishes when the
oscillation frequency goes to zero. A mirror moving slowly in the vacuum
does not appreciably squeeze it and the motional force (1) is very small at
low frequencies.

The connection of the motional force with squeezing can be analysed also in
the comoving frame. Now, the apparent input state (22) is itself squeezed
when the input is the vacuum in the laboratory frame. The motional force
thus appears as a mechanical manifestation of the fact that the mirror
scatters (with the unmodified $S-$matrix) squeezed field fluctuations.

\section*{Conclusion}

We have given the explicit expressions of the correlation functions and of
the motional force experienced by a mirror in the vacuum state. When the
mirror is irradiated by a coherent wave, the same method leads to a mean
radiation pressure and to extra fluctuations \cite{Fluctuat15}. It also
provides an extra damping force, proportional to the mirror's velocity and
to the coherent field intensity \cite{Fluctuat16}. These two results are
related through the fluctuation-dissipation theorem which would hold also
for a mirror in a thermal field.

We have only considered in this paper the one mirror problem. The situation
where two mirrors scatter the same field fluctuations seems attractive. The
mean Casimir force is well known for motionless mirrors. It is expected to
be modified when the mirrors are moving \cite{Fluctuat7,Fluctuat8}. It must
also exhibit fluctuations \cite{Fluctuat26}. These fluctuations have to be
connected with the motional effect in the same manner as in the one mirror
problem. Finally, the motional dependence of the Casimir effect is
associated to the squeezing effect due to the mirrors' motion. The formalism
developed in the present paper is applied to the two-mirrors problem in a
forthcoming paper.

\medskip
\noindent {\bf Acknowlegdements}

We thank C.Fabre, E.Giacobino and A.Heidmann for discussions.

\appendix

\section{The relations of linear response theory}

A classical modification $\delta q_{t}$ of the mirror's trajectory
corresponds to an effective perturbation (32) of the Hamiltonian. The linear
response theory \cite{Fluctuat6} provides the variation of the mean force 
\[
\left\langle \delta F(t)\right\rangle ={\int }{\rm d}t^\prime \ \chi
_{FF}^{R}(t-t^\prime )\delta q(t^\prime ) 
\]
The response function $\chi _{FF}^{R}$ (the abbreviated notation $\chi $ was
used previously) is the retarded susceptibility; it is related to the force
commutator $\xi _{FF}$ and to the correlation function $C_{FF}$ (correlation
functions are supposed stationary) 
\[
\chi _{FF}^{R}(t)=2i\theta (t)\xi _{FF}(t)\qquad \xi _{FF}(t)=\frac{
C_{FF}(t)-C_{FF}(-t)}{2\hbar }\qquad C_{FF}(t)=\left\langle
F(t)F(0)\right\rangle -\left\langle F\right\rangle ^ 2  
\]
It is also possible to define an advanced response 
\[
\chi _{FF}^{A}(t)=-2i\theta (-t)\xi _{FF}(t)=\chi _{FF}^{R}(-t) 
\]

These relations have a simple form in the spectral domain 
\begin{eqnarray*}
\left\langle \delta F[\omega ]\right\rangle  &=&\chi _{FF}^{R}[\omega
]\delta q[\omega ]\qquad \chi _{FF}^{R}[\omega ]=\widetilde{\xi }
_{FF}[\omega ]+i\xi _{FF}[\omega ] \\
\chi _{FF}^{A}[\omega ] &=&\widetilde{\xi }_{FF}[\omega ]-i\xi _{FF}[\omega
]=\chi _{FF}^{R}[\omega ]^{*}=\chi _{FF}^{R}[-\omega ]
\end{eqnarray*}
The dispersive part $\widetilde{\xi }_{FF}$ of the susceptibility functions
is obtained from the spectral density $\xi _{FF}$ through a dispersion
relation 
\[
\widetilde{\xi }_{FF}[\omega ]=\int \frac{{\rm d}\omega ^\prime }{\pi }P
\frac{\xi _{FF}[\omega ^\prime ]}{\omega ^\prime -\omega }
\]
The response function $\chi _{FF}^{R}$ (respectively $\chi _{FF}^{A}$) is
analytic (and regular) in the upper half plane $\Im \omega >0$ 
(respectively in the lower half plane $\Im \omega <0$). 
As $F$ is a Hermitean operator, $\xi _{FF}$ is a real and odd
function of $\omega $ while $\widetilde{\xi }_{FF}$ is a real and even
function of $\omega $.

The general form of the relation between the noise spectrum $C_{FF}$ and the
retarded susceptibility $\chi _{FF}^{R}$ is 
\[
\xi _{FF}(t)=\frac{C_{FF}(t)-C_{FF}(-t)}{2\hbar }=\frac{\chi
_{FF}^{R}(t)-\chi _{FF}^{R}(-t)}{2i} 
\]
which leads to equation (31) in the frequency domain 
\[
\xi _{FF}[\omega ]=\frac{C_{FF}[\omega ]-C_{FF}[-\omega ]}{2\hbar }=\frac{
\chi _{FF}^{R}[\omega ]-\chi _{FF}^{R}[-\omega ]}{2i} 
\]

\end{document}